\documentclass[superscriptaddress,reprint,pra]{revtex4-1}
\usepackage{graphicx}
\usepackage[margin=0.7in]{geometry}
\usepackage{amsmath}
\usepackage{hyperref}
\usepackage{multirow}

\let\hatOrig\hat
\renewcommand{\vec}[1]{\boldsymbol{\mathbf{#1}}}
\renewcommand{\hat}[1]{\boldsymbol{\mathbf{\hatOrig{#1}}}}
\renewcommand{\Im}{\operatorname{Im}}

\newcommand{\D}{\mathrm{d}}
\newcommand{\sub}[1]{\ensuremath{_{\textrm{#1}}}} \newcommand{\super}[1]{\ensuremath{^{\textrm{#1}}}}  \newcommand{\partialderiv}[2]{\ensuremath{\frac{\partial #1}{\partial #2}}}

\newcommand{\Kavli}{Kavli Institute for Theoretical Physics, University of California, Santa Barbara CA, USA}
\newcommand{\EPFL}{Laboratory of Nanoscience for Energy Technologies, IGM-STI, \'Ecole polytechnique f\'ed\'erale de Lausanne, Switzerland}
\newcommand{\JCAP}{Joint Center for Artificial Photosynthesis, California Institute of Technology, Pasadena CA, USA}
\newcommand{\MSC}{Materials and Process Simulation Center, California Institute of Technology, Pasadena CA, USA}
\newcommand{\Watson}{Thomas J. Watson Laboratories of Applied Physics, California Institute of Technology, Pasadena CA, USA}
\newcommand{\RPIMSE}{Department of Materials Science and Engineering, Rensselaer Polytechnic Institute, Troy, NY, USA}
\newcommand{\HarvardSEAS}{John A. Paulson School of Engineering and Applied Sciences, Harvard University, Cambridge, MA, USA}

\begin{document}

\title{Transport of hot carriers in plasmonic nanostructures}

\author{Adam S. Jermyn}\affiliation{\Kavli}
\author{Giulia Tagliabue}\affiliation{\EPFL}\affiliation{\Watson}
\author{Harry A. Atwater}\affiliation{\Watson}\affiliation{\JCAP}
\author{William A. Goddard III}\affiliation{\JCAP}\affiliation{\MSC}
\author{Prineha Narang}\email{prineha@seas.harvard.edu}\affiliation{\HarvardSEAS}
\author{Ravishankar Sundararaman}\email{sundar@rpi.edu}\affiliation{\RPIMSE}

\date{\today}

\begin{abstract}
Plasmonic hot carrier devices extract excited carriers
from metal nanostructures before equilibration, and have the
potential to surpass semiconductor light absorbers.
However their efficiencies have so far remained well below
theoretical limits, which necessitates quantitative prediction
of carrier transport and energy loss in plasmonic structures
to identify and overcome bottlenecks in carrier harvesting.
Here, we present a theoretical and computational framework,
Non-Equilibrium Scattering in Space and Energy (NESSE),
to predict the spatial evolution of carrier energy distributions
that combines the best features of phase-space (Boltzmann)
and particle-based (Monte Carlo) methods.
Within the NESSE framework, we bridge first-principles electronic structure
predictions of plasmon decay and carrier collision integrals at the atomic scale,
with electromagnetic field simulations at the nano- to mesoscale.
Finally, we apply NESSE to predict spatially-resolved energy distributions
of photo-excited carriers that impact the surface of experimentally realizable
plasmonic nanostructures at length scales ranging from tens to several
hundreds of nanometers, enabling first-principles design of hot carrier devices.
\end{abstract}

\maketitle

Surface plasmon resonances shrink optics to the nano scale,
facilitating strong focusing and localized absorption of
light \cite{Altewischer:2002tw, Gramotnev:2010pi,
Atwater:2010ys, Schuller:2010pt, Brongersma:2015fk}.
Decay of plasmons generates energetic 
electrons and holes in the material that can be exploited
for applications including photodetection, imaging and spectroscopy
\cite{Takahashi:2011ve, Wang:2011ly, HotCarrierImaging,
MappingEmission, Schuck:2013kx, Adleman:2009it, Awazu:2008nb},
photonic energy conversion, and photocatalysis \cite{Brus:2008cp, Christopher:2011ch,
Chen:2011lt, Cushing:2013zv, Frischkorn:2006py,Wu:2015dq}.
However, these applications require carriers that retain
a significant fraction of their energy absorbed from the plasmon,
which is typically two orders of magnitude larger than the thermal energy scale.
Experimentally, the energy distributions of hot carriers that critically
impact their efficiency of collection cannot be measured directly, but must instead
be inferred indirectly from optical response in pump-probe
measurements,\cite{PhysRevB.57.11334,Furube:2007lo,Harutyunyan:2015nx}
from photo-current measurements,\cite{Knight:2011ai,Zheng:2015fk}
or from redox-reaction chemical markers.\cite{Buntin:1988fx}
This critically necessitates theoretical prediction of
charge transport in metal nanostructures far from equilibrium,
which presents a major challenge for current computational methods \cite{Ziman:Principles,
LowDimensionalTransport, ValleeElectronDynamics, Tame:2013fu}.

In extremely small nano-scale systems, electron dynamics
require a full quantum mechanical treatment, and several classes
of techniques have been developed for quantum transport simulations.
In diagrammatic many-body perturbation theory, quantum transport
can be described using the non-equilibrium Greens function (NEGF)
formalism \cite{NEGF-book}, which has been applied extensively to
electron transport in molecular junctions \cite{NEGF-MolecularJunctions}.
atoms in rarefied gases \cite{NEGF-atoms},
nanoscale metal interconnects \cite{NEGF-MetalInterconnects},
and small plasmonic nanoparticles \cite{NEGF-PlasmonicSolarCell}.
Open quantum system approaches applied to photons have similarly enabled
efficient prediction of retardation and radiative effects on plasmon resonances
of nanostructures \cite{RadiativeShiftsDimers,RetardationEffectChains}.
Correspondingly, within the density-functional formalism,
time-dependent density-functional theory (TD-DFT) \cite{TDDFT} and
simplified non-adiabatic molecular dynamics (NAMD) \cite{NAMDreview}
simulations have also been used to describe electron transport
in molecules \cite{ChargeTransferTDDFT,TDDFT-QuantumTransport},
at material interfaces \cite{GoldTitaniaNAMD,CT-vdW-NAMD},
and in small plasmonic nanoparticles \cite{HotCarrierJellium,TDDFT-GoldNP}.
Both NEGF and TD-DFT methods can been applied to systems
approaching tens of nanometers in dimension using simplified
free-electron-like models.
However, in first principles simulations retaining detailed
electronic structure information, these techniques are limited by
computational complexity to at most a few hundred atoms,
corresponding to dimensions of a few nanometers.

Plasmonic nanostructures designed for harvesting hot carriers
typically range from ten to several hundred nanometers in dimensions,
well beyond dimensions where quantum transport simulations would be practical.
Additionally, with increasing dimensions, classical transport becomes
a better approximation, appropriate for hot carrier transport in these devices.
Classical transport methods include stochastic approaches that
track dynamics of individual particles, and probabilistic approaches that
describe the evolution of distribution functions.
The Boltzmann transport equation, of the latter kind, is still
computationally intensive in its most general form because it requires
tracking probability distributions in a six-dimensional phase space of
spatial and momentum degrees of freedom \cite{PhysRevB.13.4672, Pitchford2005}.
Conventional simplifications of the Boltzmann equation include
restriction of the space of allowed distribution functions,
or simplified collision integrals such as the relaxation-time
approximation \cite{PhysRev.148.766, PhysRevLett.86.2297, PhysRevB.1.2362}.
but these neglect key electronic structure details
critical in plasmonic hot carrier transport.
On the other hand, stochastic approaches such as
Monte Carlo (MC) simulations \cite{binder1986introduction}
introduce significant computational advantages for simple models
of collisions, but they become much more computationally demanding
for a complex collision model such as one based on electronic structure theory.

We have previously shown the importance of describing
hot carrier generation and transport in plasmonic materials
with full electronic structure details in the densities of states
of carriers and their matrix elements for optical transitions
and interactions with phonons \cite{NatCom,PhononAssisted}.
Neglecting spatial transport, we have combined these
calculations with simplified Boltzmann equation solutions to elucidate
non-equilibrium effects in the ultrafast spectroscopy of small
plasmonic nanoparticles \cite{TAparameters,TAanalysis}. However,
capturing the spatial variation of carrier distributions is
critical in larger and more complex plasmonic nanostructures,
where carrier generation is strongly inhomogeneous and often localized
near electromagnetic hot spots \cite{BowtieHotspots,GoldGaNcarriers}.
Simultaneously capturing electronic structure details with spatial transport
in realistic plasmonic nanostructures has so far remained a challenge.

In this \emph{Article}, we present a hybrid computational framework for
efficiently describing non-equilibrium classical charge transport
that combines the advantages of the probabilistic and stochastic approaches.
In section~\ref{sec:CompFramework}, we derive this Non-Equilibrium
Scattering in Space and Energy (NESSE) framework
as a limit of the Boltzmann equation by tracking distribution
functions indexed by number of collisions, under the assumption
of momentum randomization at each collision.
We then specialize the general NESSE approach to plasmonic
hot carriers in section~\ref{sec:PlasmonicHotCarriers},
linked with first-principles calculations of carrier generation
and scattering, as well as electromagnetic field simulations.
In section~\ref{sec:Results}, we use NESSE to predict
the spatially-resolved energy distributions of hot carriers
that reach the surface in metal nanostructures of various materials
and geometries up to several hundred nanometers in dimensions,
simultaneously accounting for electronic structure detail and nanoscale geometry.
These spatially-resolved hot carrier energy distributions are vital
to understand optical, electronic or chemical signatures of hot carriers
in experiments, and provide a direct mechanistic understanding of the transport
and energy relaxation effects which are only indirectly measurable experimentally.

\begin{figure}
\includegraphics[width=\columnwidth]{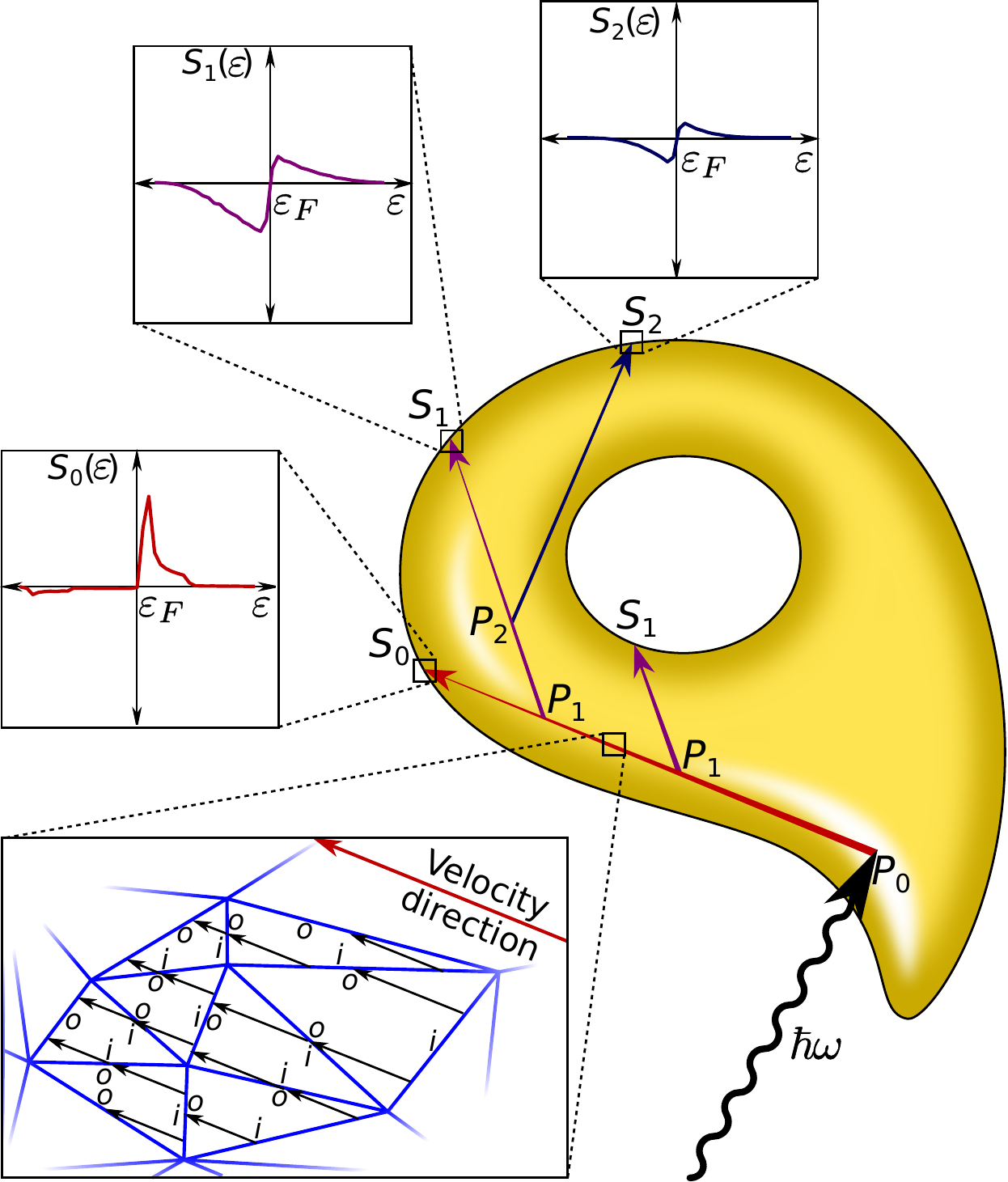}
\caption{Non-Equilibrium Scattering in Space and Energy (NESSE)
framework for evaluating non-equilibrium carrier transport in plasmonic nanostructures.
Electromagnetic field simulations and first-principles calculations of plasmon
decay determine the spatially-resolved initial carrier distribution, $P_0$.
Tracing the distributions using \emph{ab-initio} calculated mean free paths and
collision integrals determines the collected surface flux before scattering $S_0$,
and the distribution of carriers after one scattering event $P_1$.
Repeating this process yields energy-resolved carrier fluxes after
fixed numbers of scattering events $S_0$, $S_1$, $S_2$ etc. (upper insets).
Propagation of carrier distributions between scattering events is calculated efficiently
on a tetrahedral discretization of the nanostructure geometry (bottom inset).
NESSE effectively solves the linearized Boltzmann equation simultaneously 
for the spatial and energy distributions of carriers, assuming momentum
is randomized at each scattering event, in structures with arbitrary geometry
and topology, as illustrated by the arbitrarily complex shape chosen above.
\label{fig:Schematic}}
\end{figure}

\section{Computational Framework (NESSE)}
\label{sec:CompFramework}

The general goal of non-equilibrium transport calculations
is to predict the distribution of particles in phase space,
i.e. with spatial as well as momentum resolution,
accounting for sources of particles and
the various scattering mechanisms between particles.
For example, for plasmonic hot carrier devices
we need to predict the usable carrier distribution
that reaches the surface above a threshold energy,
starting from the initial distribution generated by plasmon decay
and accounting for electron-electron, electron-phonon
and surface scattering in the material.
Such devices are typically operated under constant illumination,
where carriers are being generated at a constant rate.
This results in steady-state time-independent carrier energy distributions,
while we are interested primarily in the far-from-equilibrium hot-carrier
component of these energy distributions (far from the Fermi level).
Additionally, in most cases, the number of hot carriers excited
is a small fraction of the number of electrons in the plasmonic metal,
such that hot carriers predominantly scatter against thermal carriers,
enabling a linearization of the transport equations as we discuss below.
Hence, in this work, we focus on far-from-equilibrium transport in the linearized 
steady-state limit, which is the predominant regime for hot-carrier solar energy harvesting.

In steady state, the general transport problem is described by the time-independent Boltzmann equation,
\begin{equation}
\vec{v}_s \cdot \nabla f(s,\vec{r}) = P_0(s,\vec{r}) + \Gamma_s[f],
\label{eqn:BoltzmannSS}
\end{equation}
in terms of the spatially-varying state occupation $f(s,\vec{r})$.
The abstract state label $s$ includes all degrees of freedom at a
given point in space, which is just momentum $\vec{p}$ in the classical case.
For electrons in a material, $s$ combines crystal momentum
$\vec{k}$ in the Brillouin zone with a band index $n$.

The term on the left side of (\ref{eqn:BoltzmannSS}) accounts
for drift of particles in state $s$ with velocity $\vec{v}_s$.
The first term on the right side, $P_0$, accounts for particle generation,
while the second term, the collision integral $\Gamma$ accounts for scattering.
See section~\ref{sec:PlasmonicHotCarriers} for a complete specification
of these terms for the plasmonic hot carrier example starting
from the electronic structure of the material.

Once the source term and collision integrals have been defined,
the Boltzmann equation is fully specified and can, in principle, be solved.
However, this deceptively simple-looking equation is a nonlinear integro-differential
equation in six dimensions, differential in the three spatial dimensions $\vec{r}$
and integral (non-local) in the three momentum dimensions in $\vec{k}$ within $s$,
which makes it extremely expensive computationally.
The remainder of this section develops a practical approximation to
this equation that is suited for analyzing hot carrier transport and related scenarios.

The first substantial simplification is linearization of the collision integral,
which is possible whenever the particles in which we are interested in scatter predominantly
against a background of particles whose distribution is fixed or already known.
For plasmonic hot carriers, this is the case in the low intensity regime,
where the number of excited far-from-equilibrium carriers
is small compared to the background of equilibrium carriers.
In this regime, hot carriers scatter predominantly against equilibrium carriers,
and phonons remain approximately in equilibrium at the ambient temperature $T_0$.
We can then separate $f(s,\vec{r}) = f_0(\varepsilon_s,T_0) + \phi(s,\vec{r})$,
where the first term is the equilibrium (Fermi) distribution,
and the second term is the deviation from equilibrium.
Substituting this into (\ref{eqn:BoltzmannSS}), Taylor expanding
the collision integral about the equilibrium Fermi distribution $f_0$,
and dropping terms at second order and higher in $\phi(s,\vec{r})$ yields
the linearized steady-state Boltzmann equation,
\begin{equation}
\vec{v}_s \cdot \nabla \phi(s,\vec{r}) = P_0(s,\vec{r}) + \sum_{s'} C_{ss'} \phi(s',\vec{r}),
\label{eqn:BoltzmannLinearized}
\end{equation}
where the `collision matrix' $C$ arises from the first order term
in the Taylor expansion of the collision integral,
\begin{equation}
C_{ss'} = \left. \partialderiv{\Gamma_s[f]}{f(s')} \right|_{f=f_0}.
\label{eqn:CollisionMatrix}
\end{equation}
The zeroth order terms in this equation, which correspond to the equilibrium configuration, cancel by definition.
Above and henceforth, any sum over a state index $s$ is understood to imply 
integration over the continuous $\vec{k}$ degrees of freedom contained within.

Importantly, while the collision integral $\Gamma[f(s,\vec{r})]$ varies spatially
due to the spatial dependence of $f(s,\vec{r})$, the collision matrix does not.
In the hot carrier case, linearization is only valid for energies several $k_BT$ away from
the Fermi level because electron-electron scattering produces more low-energy carriers,
eventually affecting all electrons near the Fermi energy in the material,
regardless of the incident intensity and initial number of carriers.
Assuming linearity (inherently an approximation) is adequate for analyzing plasmonic hot carrier devices,
but retaining the nonlinearity is important for describing the thermalized regime
and, of course, for high-intensity regimes explored in ultrafast spectroscopy,
which we have analyzed in detail elsewhere (neglecting spatial dependence
instead in that case) \cite{TAparameters,TAanalysis}.

The linearized steady-state Boltzmann equation (\ref{eqn:BoltzmannLinearized})
remains extremely challenging to solve since it still requires keeping track
of a six-dimensional distribution function.
In order to address this issue, we rearrange the equation to separate
the distribution functions by the number of scattering events.
First, we separate the diagonal terms of the scattering matrix,
which correspond to the state inverse-lifetimes $\tau_s^{-1}$,
to write
\begin{equation}
C_{ss'} \equiv -\tau_s^{-1} \delta_{ss'} + M_{ss'}
\label{eqn:MixingMatrix}
\end{equation}
and thereby define the `mixing matrix' $M_{ss'}$.
Intuitively, $M_{ss'}$ specifies the rate of generating carriers in
state $s$ due to scattering of a carrier in state $s'$.
Substituting (\ref{eqn:MixingMatrix}) into (\ref{eqn:BoltzmannLinearized})
and rearranging yields
\begin{equation}
\left(\tau_{s}^{-1} + \vec{v}_s\cdot\nabla\right) \phi(s,\vec{r}) = P_0(s,\vec{r}) + \sum_{s'}  M_{ss'} \phi(s',\vec{r}).
\label{eqn:BoltzmannLinearized2}
\end{equation}
Now substitute $\phi = \phi_0 + \phi_1 + \phi_2 + \cdots$ above,
where $\phi_n$ collects contributions at $n\super{th}$ order in $M$,
and collect terms by order in $M$.
This leads to the recurrence relations
\begin{flalign}
\left(\tau_{s}^{-1} + \vec{v}_s\cdot\nabla\right) \phi_n(s,\vec{r}) &= P_n(s,\vec{r}) \label{eqn:phi_n}\\
P_{n+1}(s,\vec{r}) &= \sum_{s'} M_{ss'} \phi_n(s',\vec{r}). \label{eqn:Pn}
\end{flalign}
for $n \ge 0$, with the initial point $P_0$ given as before by (\ref{eqn:P0}).

Figure~\ref{fig:Schematic} illustrates this formulation of Boltzmann transport.
Optical absorption produces carriers at rate $P_0$.
Solving (\ref{eqn:phi_n}) yields unscattered hot carrier distribution $\phi_0$.
Applying the mixing matrix to this in (\ref{eqn:Pn}) then calculates
the rate of generating carriers due to the first scattering event, $P_1$.
The process then repeats to evaluate carrier distributions after one scattering, $\phi_1$,
generation rate due to the second scattering event, $P_2$, and so on.
For each $n$, the carrier flux reaching the surface of the nanostructure
after $n$ scattering events can be evaluated as $S_n(s,\vec{r}) = (\vec{v}_s\cdot\hat{a})
\phi_n(s,\vec{r})$, at point $\vec{r}$ with surface normal unit vector $\hat{a}$.

The bottleneck at this stage is keeping track of states $s$
with wave vector and band indices. In particular, the mixing matrix $M_{ss'}$
is dense with non-zero elements for all pairs of $s$ and $s'$
(i.e. total six dimensions with $\vec{k}$ and $\vec{k}'$),
which makes it computationally impractical to store and use.
Our final simplification is to assume that each scattering event randomizes the momentum direction,
which is an excellent approximation for hot carriers in plasmonic metals (see e.g. Fig. (3d) of Ref.~\citenum{GraphiteHotCarriers}).
We can now define carrier distributions with respect to energy
by summing over all states that yield a given energy,
\begin{equation}
X(\varepsilon,\vec{r}) \equiv \sum_s \delta(\varepsilon - \varepsilon_s) X(s,\vec{r})
\end{equation}
for each distribution function $X = P_n$, $\phi_n$ and $S_n$.
We solve for $\phi_0(s,\vec{r})$ using explicit states as in (\ref{eqn:phi_n}), since this
is before first scattering, but from that point onward, we only work with energy distributions.
The randomized-momentum equations after obtaining $\phi_0(\varepsilon,\vec{r})$ take the form
\begin{flalign}
P_{n+1}(\varepsilon,\vec{r}) &= \int \D\varepsilon' 
	M(\varepsilon,\varepsilon') \phi_n(\varepsilon',\vec{r}) \label{eqn:PnE}\\
\phi_n(\varepsilon,\vec{r}) &= \int \frac{\D\hat{v}}{4\pi}
	\left(\bar{\tau^{-1}}(E) + \bar{v}(E)\hat{v}\cdot\nabla\right)^{-1}
	P_n(\varepsilon,\vec{r}), \label{eqn:phi_nE}
\end{flalign}
where $\bar{\tau^{-1}}(\varepsilon)$ and $\bar{v}(\varepsilon)$
are the average inverse lifetime and speed of carriers with energy $\varepsilon$,
and the mixing-matrix in energy is
\begin{equation}
M(\varepsilon,\varepsilon') \equiv \sum_{ss'}
	\delta(\varepsilon-\varepsilon_s) M_{ss'}
	\delta(\varepsilon_{s'}-\varepsilon').
\label{eqn:MixingMatrixE}
\end{equation}
Note that the averages above are defined by $\bar{x}(\varepsilon)
\equiv \sum_s \delta(\varepsilon-\varepsilon_s) x_s / g(\varepsilon)$
for each quantity $x$, where $g(\varepsilon) = \sum_s \delta(\varepsilon-\varepsilon_s)$
is the density of states.
In numerical simulations for plasmonic hot carriers, we discretize carrier energy on a uniform grid
extending from $\hbar\omega$ below to $\hbar\omega$ above the Fermi energy,
and perform the integral over $\hat{v}$ in (\ref{eqn:phi_nE}) by Monte-Carlo sampling.

The final remaining ingredient is the spatial propagation of distributions
at a given $s$ (for $n=0$) or a given pair of $E$ and $\vec{v}$ (for $n \ge 1$),
both of which require solution of a differential equation of the form
\begin{equation}
\left(\tau^{-1} + \vec{v}\cdot\nabla\right) \phi(\vec{r}) = P(\vec{r}).
\end{equation}
Importantly, this is an ordinary differential equation, which can be solved
very efficiently by appropriate choice of coordinate system.
Without loss of generality, pick the $z$-axis along $\vec{v}$
to get $\phi(z)/\tau + v(\D\phi/\D z) = P(z)$ for each $x,y$,
which has a simple solution of the form
\begin{equation}
\phi(z) = \phi(z_0) e^{-(z-z_0)/(v\tau)} + \int_{z_0}^z \D z' \frac{P(z')}{v} e^{-(z-z')/(v\tau)}.
\label{eqn:zPropagation}
\end{equation}
For a general three-dimensional geometry discretized using a tetrahedral mesh,
we apply this solution in each tetrahedron, effectively evolving the
distribution functions from the input faces `$i$' to the output faces `$o$',
as shown in the bottom inset of Figure~\ref{fig:Schematic}.

In particular, we store $P(\vec{r})$ and $\phi(\vec{r})$ on the vertices
with (3D) linear interpolation in the interior of each tetrahedron,
and store fluxes $S(\vec{r}) = (\vec{v}\cdot\hat{a})\phi(\vec{r})$
on the vertices of every triangular face (with unit normal $\hat{a}$)
with (2D) linear interpolation in the interior of each face.
The solution within each tetrahedron can then be expressed as
matrices yielding the output $\phi$ on the volume and $S$ on `$o$' faces,
given the input $P$ on the volume and $S$ on the `$i$' faces,
where the matrix elements can be calculated by integrating
(\ref{eqn:zPropagation}) against the face and volume interpolants.
Then the solution for the whole mesh starts by applying this solution
to all tetrahedra whose `$i$' faces are exclusively incoming faces of
the structure (so that their $S$ input is already known).
This determines $S$ on `$o$' faces for all these tetrahedra, which makes
the $S$ on `$i$' faces known for a new set of tetrahedra.
The solution can then be applied to these tetrahedra,
and the process repeated till all tetrahedra are exhausted
and $S$ on the outgoing surface of the overall structure is determined.
(See bottom inset of Figure~\ref{fig:Schematic}.)

For the propagation at $n=0$, we apply the above scheme for several
electron and hole velocities and energies, obtained from a Monte Carlo
sampling of the Brillouin zone integrals in (\ref{eqn:ImEps}).
At this stage, the only source term is $P_0$, distributed on the volume,
while the output is $\phi_0$ on the volume and $S_0$ at the surface.
In subsequent stages $n \ge 1$, we apply the above scheme with a Monte Carlo sampling
as described.
For these stages, an additional surface source term is possible due to reflection
of carriers at the surface, adding $S\super{(in)}_{n+1}(\varepsilon,\vec{r})
= \alpha(\varepsilon,\vec{r}) S_{n}(\varepsilon,\vec{r})$ to the propagation scheme,
where $\alpha(\varepsilon,\vec{r})$ is the energy- and surface-dependent reflection fraction.
Here $\alpha(\varepsilon,\vec{r})$ should depend on the material outside the plasmonic metal,
varying from 0 for total internal reflection (eg. below Schottky barriers)
to 1 for perfect injection (an unattainable upper bound).
In this work we explore the effect of arbitrary reflection fractions on
the collected carrier distributions, but for specific experimental designs
we can incorporate appropriate models of carrier injection across interfaces.

\section{Plasmonic hot carrier transport}
\label{sec:PlasmonicHotCarriers}

The general NESSE framework for transport developed above
in section~\ref{sec:CompFramework} requires two quantities that
must be specified for a given problem: the source term $P_0$
and the collision integral $\Gamma$.
For plasmonic hot carrier transport, the source term i.e. the
spatially-resolved initial carrier distribution, can be evaluated as
\begin{equation}
P_0(s,\vec{r}) = \frac{1}{2\pi \hbar} \vec{E}^\ast(\vec{r}) \cdot \Im\bar{\epsilon}(\omega,s) \cdot \vec{E}(\vec{r}),
\label{eqn:P0}
\end{equation}
where $P_0(s,\vec{r})$ is the rate of generation of carriers per
unit volume at location $\vec{r}$ and at bulk state index $s = \vec{k}n$,
that combines crystal momentum $\vec{k}$ and band $n$.
Here, $\Im\bar{\epsilon}(\omega,s)$ is the imaginary dielectric tensor
at incident frequency $\omega$ histogrammed by carrier state $s$,
and $\vec{E}(\vec{r})$ is the electric field distribution in the material.
As above, in the steady-state problem, the field and carrier distributions are all time-independent.

Above, we make two approximations.
By resolving the distribution in both space and momentum ($\vec{k}$
contained within index $s$), we are making a semi-classical
approximation that precludes quantum effects in the transport, but retains
the bulk electronic structure of the material ($s$ indexes bulk states).
Additionally, we assume locality in that photons are absorbed
in the material spatially distributed by the field intensity,
and that they then produce carriers in the same location.
Both these approximations are applicable when the structures
are much larger than the nonlocality and coherence length scales
(at the nanometer scale for plasmonic metals at room temperature),
which will be the case for typical plasmonic metal nanostructures
with dimensions of at least several nanometers.

For calculating (\ref{eqn:P0}), the field distribution $\vec{E}(\vec{r})$ in a
plasmonic nanostructure of interest can be readily evaluated using any standard
finite-element method (FEM) or finite-difference time-domain (FDTD) simulation tool.
The carrier-resolved imaginary dielectric tensor $\Im\bar{\epsilon}(\omega,s)$
is obtained using our previously established \emph{ab initio} method \cite{NatCom,PhononAssisted,TAparameters,TAanalysis},
\begin{widetext}\begin{multline}
\vec{\lambda}^\ast\cdot\Im\bar{\epsilon}(\omega,s)\cdot\vec{\lambda}
= \frac{4\pi^2 e^2}{m_e^2\omega^2} \int\sub{BZ} \frac{d\vec{k}}{(2\pi)^3} \sum_{n'n}
	(\delta(s,\vec{k}n') - \delta(s,\vec{k}n))
	(f_{\vec{k}n} - f_{\vec{k}n'}) \delta(\varepsilon_{\vec{k}n'} - \varepsilon_{\vec{k}n} - \hbar\omega)
	\left| \vec{\lambda} \cdot \langle\vec{p}\rangle^{\vec{k}}_{n'n} \right|^2\\
+ \frac{4\pi^2 e^2}{m_e^2\omega^2}
	\int\sub{BZ} \frac{d\vec{k}'d\vec{k}}{(2\pi)^6} \sum_{n'n\alpha\pm}
	(\delta(s,\vec{k}'n') - \delta(s,\vec{k}n))
	(f_{\vec{k}n} - f_{\vec{k}'n'}) \left( n_{\vec{k}'-\vec{k},\alpha} + \frac{1}{2} \mp \frac{1}{2}\right)\\
\times \delta(\varepsilon_{\vec{k}'n'} - \varepsilon_{\vec{k}n} - \hbar\omega \mp \hbar\omega_{\vec{k}'-\vec{k},\alpha})
	\left| \vec{\lambda} \cdot \sum_{n_1} \left(
	\frac{ g^{\vec{k}'-\vec{k},\alpha}_{\vec{k}'n',\vec{k}n_1} \langle\vec{p}\rangle^{\vec{k}}_{n_1n} }
		{ \varepsilon_{\vec{k}n_1} - \varepsilon_{\vec{k}n} - \hbar\omega + i\eta} +
	\frac{ \langle\vec{p}\rangle^{\vec{k}'}_{n'n_1} g^{\vec{k}'-\vec{k},\alpha}_{\vec{k}'n_1,\vec{k}n} }
		{ \varepsilon_{\vec{k}'n_1} - \varepsilon_{\vec{k}n} \mp \hbar \omega_{\vec{k}'-\vec{k},\alpha} + i\eta }
\right) \right|^2,
\label{eqn:ImEps}
\end{multline}\end{widetext}
where the first and second terms capture the contributions of direct
and phonon-assisted transitions respectively, and $\vec{\lambda}$ is
an arbitrary test vector to sample the tensorial components.
Briefly, $\varepsilon_{\vec{k}n}$ and $f_{\vec{k}n}$ are electron energies
and Fermi occupations indexed by wave vector $\vec{k}$ and band $n$,
$\hbar\omega_{\vec{q}\alpha}$ and $n_{\vec{q}\alpha}$ are phonon energies
and Bose occupations indexed by wave vector $\vec{q}\equiv\vec{k}'-\vec{k}$
and polarization $\alpha$, and $\langle p\rangle$ and $g$ are respectively
the momentum matrix elements for electron-light interactions and
the electron-phonon matrix elements, all of which we calculate
\emph{ab initio} using density-functional theory.
See Ref.~\citenum{PhononAssisted} for a detailed discussion of
the above terms and the computational details in evaluating them.
The only modification is the first factor containing $\delta(s,\vec{k}n)$,
a Kronecker $\delta$ that selects the combined state index $s$
that corresponds to wavevector $\vec{k}$ and band $n$.
This histograms the contributions by carrier state: the positive
terms for final states in the transitions correspond to electrons,
while the negative terms for the initial states correspond to holes.

We have previously shown for plasmonic metals that the
DFT-calculated band structure is in excellent agreement with
$GW$ calculations and ARPES measurements \cite{NatCom,QSGW-Au},
and that the electron-phonon and optical matrix elements
result in accurate prediction of resistivity and dielectric
functions in comparison to experiment \cite{PhononAssisted}.
However, note that the NESSE framework is completely general,
and the specific electronic structure choices we make here can
be systematically improved upon using TD-DFT\cite{TDDFT}
or many-body $GW$ perturbation theory \cite{GW},
if necessary for other materials.

For transport of hot carriers in plasmonic nanostructures,
the collision integral, $\Gamma_s[f]$, must account for scattering
between electrons as well as scattering of electrons against phonons.
We can evaluate this using \emph{ab initio} band structures and matrix elements as
\begin{widetext}\begin{multline}
\Gamma_{\vec{k}n}[f]
= \frac{2}{\hbar} \int\sub{BZ} \frac{d\vec{k}'}{(2\pi)^3} \sum_{n'} \sum_{\vec{G}\vec{G}'}
	\left( f_{\vec{k}'n'} - f_{\vec{k}n} \right)
	\tilde{\rho}_{\vec{k}'n',\vec{k}n}(\vec{G})
	\tilde{\rho}_{\vec{k}'n',\vec{k}n}^\ast(\vec{G}')
	\frac{4\pi e^2}{\left| \vec{k}'-\vec{k}+\vec{G} \right|^2}
	\Im\left[ \epsilon^{-1}_{\vec{G}\vec{G}'}(\vec{k}'-\vec{k},\varepsilon_{\vec{k}n}-\varepsilon_{\vec{k}'n'}) \right]\\
+ \frac{2\pi}{\hbar} \int\sub{BZ} \frac{\Omega d\vec{k}'}{(2\pi)^3} \sum_{n'\alpha\pm}
	\left( f_{\vec{k}'n'} - f_{\vec{k}n} \right)
	\left( n_{\vec{k}'-\vec{k},\alpha} + \frac{1}{2} \mp \frac{1}{2} \right)
	\delta(\varepsilon_{\vec{k}'n'} - \varepsilon_{\vec{k}n} \mp \hbar\omega_{\vec{k}'-\vec{k},\alpha})
	\left| g^{\vec{k}'-\vec{k},\alpha}_{\vec{k}'n',\vec{k}n} \right|^2,
\label{eqn:CollisionIntegral}
\end{multline}\end{widetext}
where the first and second terms account for electron-electron and electron-phonon scattering respectively.
Briefly, the new quantities here are $\tilde{\rho}$, the density corresponding to
the product of a pair of electronic wavefunctions in reciprocal space where $\vec{G}$
are reciprocal lattice vectors, and $\epsilon^{-1}_{\vec{G}\vec{G}'}$, the frequency-dependent
inverse dielectric matrix (full nonlocal response) evaluated within the random-phase approximation.
(This is closely-related to a quasiparticle linewidth calculation
in many-body perturbation theory within the $G_0W_0$ approximation.)
See the discussion of the corresponding electron linewidth
contributions in Ref.~\citenum{PhononAssisted} for further details.
The only differences here are a factor of $2/\hbar$ to convert
from linewidth to scattering rate, and a trivial generalization
from Fermi distributions to arbitrary occupations $f$.

\section{Computational details}

We use density-functional theory calculations of electrons, phonons and their matrix elements
in the open-source JDFTx software\cite{JDFTx} to evaluate the generated carrier distributions
(\ref{eqn:ImEps}) and collision integrals in the mixing matrix form (\ref{eqn:CollisionIntegral},
\ref{eqn:CollisionMatrix}, \ref{eqn:MixingMatrix}, \ref{eqn:MixingMatrixE}).
See Ref.~\citenum{PhononAssisted} for a complete specification of the electronic structure details.

In NESSE calculations of hot carrier transport, we use $\sim 10^5$
random pairs of electron and hole states that conserve energy with the incident photons
of energy $\hbar\omega$ for the $n=0$ step (prior to the first scattering event).
For subsequent steps, we use a uniform energy grid with resolution $\sim 0.1$~eV
and use $\sim 10^4$ random samples of energy and velocity direction.
We perform the transport solution on the same tetrahedral mesh as the electromagnetic simulation.

The solution scheme detailed above scales linearly with the number of tetrahedra
and achieves a typical throughput of $1-2 \times 10^6$ tetrahedra/second per CPU core
(timed on NERSC Edison and Cori), and parallelizes linearly over the velocity and energy (state) samples.
For typical geometries requiring $\sim 10^5$ tetrahedra in the plasmonic metal,
our solution scheme with the chosen number of samples therefore takes $\sim 10^2$ seconds
on 100 cores for the $n=0$ step, and $\sim 10$ seconds for each subsequent $n$.
Note that NESSE presents significant advantages over Monte Carlo simulations
of individual carriers, because we completely avoid ray-tetrahedron intersections
and obtain solutions for the entire mesh together for a given electronic state (energy \& velocity).

For electromagnetic (EM) simulations, we use the commercial software COMSOL, based on the finite element method.
In all cases, we use the Wave Optics package, solving the EM wave equations in the frequency domain.
Furthermore, we employ the scattering field formulation, analytically defining the incident
(background) electric field with the desired polarization and computing the scattered field.
The simulated structure is placed at the center of a spherical domain and is enclosed by a perfectly matched layer.
We use a tetrahedral mesh which is highly refined inside and around the structure of interest.
Both the enclosing domain dimensions and the mesh refinement have been tested to ensure
parameter independence of the results (both EM and transport, since they share the same mesh).

\section{Results and Discussion}
\label{sec:Results}
\begin{figure}
\includegraphics[width=\columnwidth]{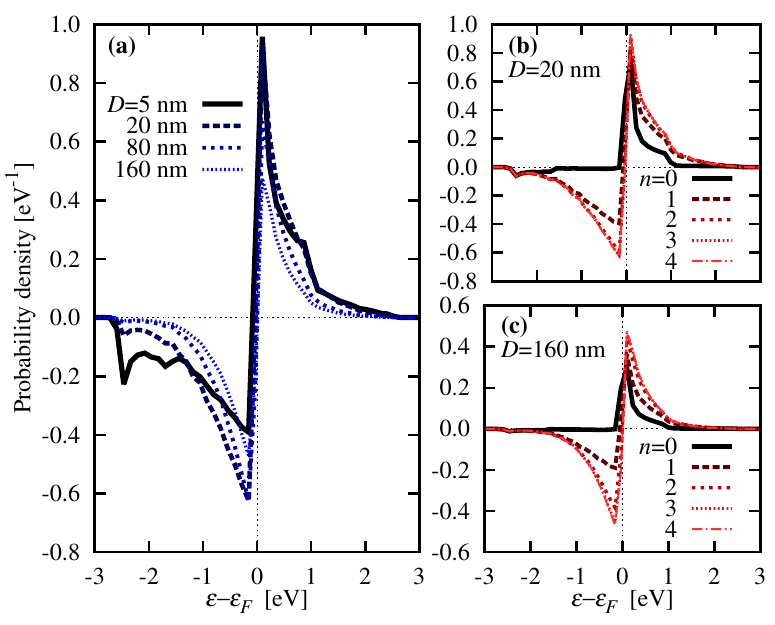}
\vspace{-0.2in}\caption{(a) Hot carrier energy distribution reaching the surface
of spherical gold nanoparticles of various diameters $D$,
normalized to one absorbed photon ($\lambda = 490$~nm).
This distribution is the perturbation from equilibrium:
negative probabilities here correspond to holes.
Note increasing thermalization with increasing $D$.
(b,c) Cumulative carrier distributions that reach after
scattering $\le n$ times in 20 and 160 nm nanoparticles.
Note that holes predominantly reach after 1 scattering event,
and the contributions of successive scattering events
diminishes rapidly (at a faster rate for smaller particles).
\label{fig:ScatterSizeDep}}
\end{figure}

Our NESSE framework resolves carrier dynamics spatially, energetically, and as a function of the number of times they scatter.
Therefore it provides fine-grained information on the physics of transport of hot carriers.
To explore this data we begin with Figure~\ref{fig:ScatterSizeDep}(a),
which shows the carrier distribution that reaches the surface of a spherical gold nanoparticle as a function of energy and the size of the particle.
Smaller nanoparticles collect a larger fraction of their carriers at high energy.
This is because these carriers have had less distance (and therefore fewer opportunities) to scatter.

To begin, we examine the carrier distribution in energy and scattering count.
Figure~\ref{fig:ScatterSizeDep} shows the collected carrier distributions, both
(a) total and (b-c) by scattering event, in spherical gold nanoparticles of various diameters.
As expected, the smaller particles collect more carriers at low scattering counts.
In addition, even at a given scattering count the smaller particle
collects carriers of higher energy than the larger particle, because the
mean free path of carriers decreases with increasing energy \cite{PhononAssisted}.

Broadly speaking then there are two primary effects which vary with particle size:
the first is a change in regime; small particles collect carriers nearly ballistically,
whereas large particles collect them semi-diffusively.
The second is a change in energy scale; large particles preferentially collect
low-energy carriers as high-energy carriers scatter more readily.
Note that small nanoparticles will additionally generate carriers by
intraband transitions due to the nanoscale field geometry (Landau damping).
However, phonon-assisted transitions dominate over this effect in particles
larger than about 40~nm \cite{PhononAssisted}, which is the regime in which
transport effects are important anyway. We therefore focus on these larger
nanostructures and do not explicitly include geometry-assisted carrier
excitations in the results presented below.

Figures~\ref{fig:ScatterSizeDep}(b-c) reveal another key feature:
the asymmetry between electron and hole scattering.
High-energy holes in gold are located in the $d$ bands
and have a much smaller mean free path than electrons
of comparable energy because of lower band velocities \cite{PhononAssisted}.
As a result almost no holes are collected before the first scattering event.
Collection then peaks after the first scattering and
subsequently diminishes rapidly as the holes thermalize.
The implication of this result for hole-driven solid-state
and photochemical systems is that structures should be designed
either below the high-energy hole mean free path length scale
($\sim 1-3$~nm) to collect them before the first scattering event,
or around the intermediate-energy hole mean free path scale
($\sim 10$~nm) to exploit the once-scattered holes.

Finally, Figures~\ref{fig:ScatterSizeDep}(b-c) show that with increasing
numbers of scattering events, the carrier energy distributions approach
symmetric electron and hole perturbation close to the Fermi level,
corresponding to an increase of electron temperature as expected.
Notice that it takes only 3-4 scattering events to approach this limit,
underscoring the importance of designing hot carrier devices in
the sub-to-few mean free path scale.

The results discussed above so far assumed perfect collection at the surface.
In realistic metal-semiconductor interfaces, injection of carriers
from the metal to the semiconductor is possible only when the carrier energy
exceeds the Schottky barrier height, and when the carrier momentum
tangential to the interface is conserved across it \cite{InternalPhotoemission}.
This energy and momentum-dependent injection probability limits
carrier collection efficiency \cite{Leenheer:2014sw},
and is sensitive to the energy-momentum dispersion relations of both
the metal and the semiconductor, their energy level alignment at the interface,
as well as to the roughness of the interface \cite{HotCarrierReflection}.
The NESSE framework contains information regarding the momentum distributions
impinging on the surface for each scattering count $n$, and can 
readily be coupled to a detailed model of the injection probability.

\begin{figure}
\includegraphics[width=\columnwidth]{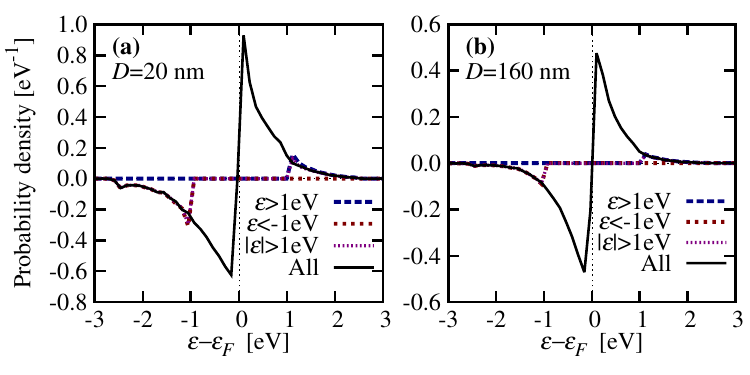}
\vspace{-0.2in}\caption{Dependence of hot carrier energy distribution reaching the
surface on what carrier energies are allowed to pass through,
for particles of diameter (a) 20 nm and (b) 160 nm.
When only electrons with $\varepsilon > 1$~eV pass through,
reflected holes slightly enhance the number of these electrons,
and vice versa, when only holes with $\varepsilon < -1$~eV pass through.
However this is a small effect, especially for larger dimensions,
so it is safe to assume that the carrier distribution reaching the surface
is independent of the selection rules for hot carrier extraction at the surface.
\label{fig:BounceDep}}
\end{figure}

Here, we focus on the effect of carrier injection on carrier transport
within the metal, and therefore adopt simplified injection probability models
that would introduce the \emph{maximum effect on carrier transport}.
Figure~\ref{fig:BounceDep} examines the carrier distribution collected
at the surface of a gold nanoparticle with varying surface collection properties.
We compare four distinct scenarios here, one in which all carriers are collected (ideal)
and three in which carriers are only collected above a specific energy.
Note that the carriers that are not collected are assumed to reflect back
into the material, where they can undergo further scattering processes.

These scenarios suggest that, by and large, fractional reflection of carriers
at one energy has the primary effect of reducing carrier collection at that energy.
The secondary effect of reintroducing the reflected carriers into the material
only minimally enhances collection at other energies.
Moreover this enhancement is limited to energies just above the collection threshold,
and noticeable only for smaller particles where the reflected carriers are likely
to reach another surface prior to additional scattering.
This observation has an important implication both for experimental design
and hot carrier device simulation: the available hot carrier flux at
the surface is mostly independent of the surface collection property,
cleanly separating the geometric design of the metal structure
with the material design of the metal-collector interface.

\begin{figure}
\includegraphics[width=\columnwidth]{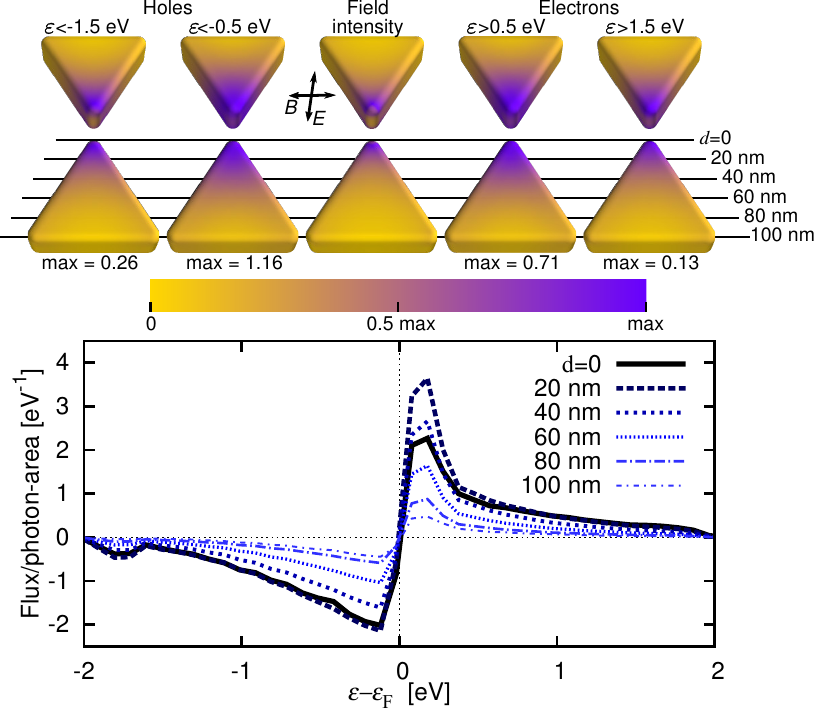}
\caption{Spatially-resolved hot carrier energy distribution reaching
the surface of a gold bowtie nanoantenna illuminated on resonance (650~nm).
The top panels compare the spatial distribution of the electric field
with fluxes of electrons or holes with energy further than
0.5 or 1.5~eV from the Fermi level, while the lower panels show the
average energy-resolved carrier flux in various spatial slices.
The fluxes are normalized per absorbed photon per unit area of the surface.
Higher energy carriers decay rapidly and are localized to the electromagnetic
`hot spots', whereas lower energy carriers can reach further away.
\label{fig:Bowtie}}
\end{figure}

Having examined the dependence on energy and scattering properties in
spherical particles where spatial variations are less important,
we now turn to complex structures with high spatial inhomogeneity
and exploit the full power of the NESSE framework.
Figure~\ref{fig:Bowtie} shows carrier distributions resolved
in energy and space for a gold bowtie nanoantenna
(100~nm equilateral triangles with 40~nm thickness,
10~nm corner radius and a 30~nm gap), illuminated at
normal incidence with the electric field direction
along the gap and at a resonant wavelength of 650~nm.
For this structure and illumination, the field intensity (central top panel),
and hence the initial distribution of generated carriers,
are sharply localized near the gap.
The remaining top panels show the carrier fluxes of holes (left)
and electrons (right) above different cutoff energies,
normalized per absorbed photon per unit total surface area.
Note that these normalized fluxes are dimensionless, but they are not
probabilities; they can exceed one both because they are a ratio of
carrier flux at one point to the \emph{average} absorbed photon flux,
and because each electron-electron scattering event produces multiple
lower energy hot carriers.

The collected carriers localize quite strongly near the field maximum,
indicating that field enhancement remains a significant factor
even after transport processes are accounted for.
Importantly, however, the extent of localization of collected carriers
is strongly dependent on the energy of collected carriers.
With a higher energy threshold for collection, the overall
carrier flux diminishes and becomes more strongly localized
towards the high-field gap region.
This is also shown in the carrier energy distributions
reaching the surface at various distances from the gap
region in the bottom panel of Figure~\ref{fig:Bowtie}.
Low-energy carriers can reach further from the
high-field `hot spots' both because of their higher
mean free path and because they can be generated
by scattering of higher energy carriers.

Overall these results show that both the electromagnetic field
distribution and the carrier scattering properties play a vital
role in shaping the carrier distributions that reach the surface.
Predictions solely based on field intensity will overstate
carrier localization, particularly at lower energies,
while predictions made without accounting for the field
distribution will completely miss the spatial inhomogeneity.
Experimentally, this spatial inhomogeneity on the tens to hundreds
of nanometers is vital to understand hot carrier imaging, photodetection,
photovoltaic and photochemical energy conversion, which all involve
carrier collection into a semiconductor or molecule.\cite{BowtieHotspots,GoldGaNcarriers}
However, optical probes of the carrier response (which we can
predict quantitatively using our first-principles framework as
shown previously \cite{TAanalysis}) cannot sense this inhomogeneity
due to the diffraction limit, and instead measure a spatially-averaged result.
The hot carrier distribution that we predict here is critical to
understand experiments where hot carrier transport matters, precisely
because it is not possible to measure these distributions directly.

\begin{figure}
\includegraphics[width=\columnwidth]{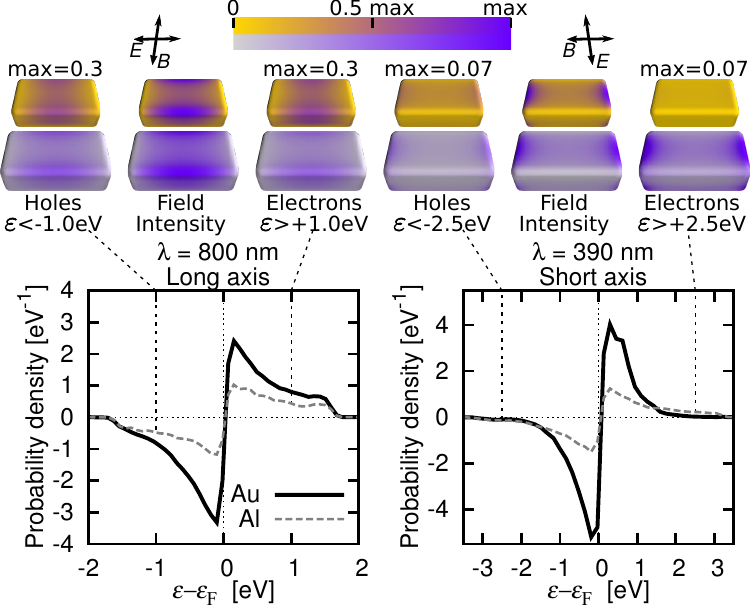}
\caption{Comparison of spatially- and energy-resolved hot carrier fluxes
reaching the surface of a gold (upper) and aluminum (lower) rectangular nanorod
illuminated at 800 nm wavelength for the long axis (left) and
390~nm wavelength for the short axis (right).
The top panels compare the spatial distributions of carrier flux with
energy greater than 1.5eV (left) or 2.5eV (right) away from the Fermi level,
while the lower panels show the spatially-averaged energy distributions.
Fluxes are normalized per photon absorbed per unit area of the surface.
Aluminum is less efficient at collecting low energy carriers due to
a shorter electron-phonon scattering mean free path, but enhances
high-energy electron collection compared to gold because of increased
generation of (and comparable mean free paths for) high-energy electrons.
\label{fig:Nanorod}}
\end{figure}

Finally, we investigate the influence of material choice on the carrier distribution,
and in particular, due to the increasing interest in aluminum plasmonics, compare
carrier distributions in gold and aluminum structures with similar geometries.
In particular, we pick the rectangular nano-antenna geometry shown in Figure~\ref{fig:Nanorod},
with long-axis length 140~nm, short-axis length 80~nm, height 40~nm and 10~nm corner radius.
At normal incidence, the resonant absorption frequency depends both on the material
and whether the polarization (electric field) direction is along the long or short axis.
In particular, we find broad absorption resonances centered at 600~nm and 800~nm
respectively for aluminum and gold for the long-axis polarization,
and at 390~nm and 610~nm respectively for the short-axis polarization.
In order to compare carrier dynamics keeping all other parameters similar,
we pick a common long-axis illumination wavelength of 800~nm and a
short-axis illumination wavelength of 390~nm, which covers the greatest
range of photon energies and exhibits strong absorption in both materials.
Since we care about the collected carriers \emph{per absorbed photon},
minor differences in the absolute absorption cross section are irrelevant.

The top panels of Figure\ \ref{fig:Nanorod} show the spatially-resolved
carrier fluxes above a threshold energy in both materials and for both illuminations.
As before, the carriers are less localized than the field intensity
and the spatial extent increases with decreasing energy due to the higher
mean free path and secondary-scattered contributions at lower energies.
This effect is comparable in the two materials.
Now, compare the probabilities of carrier collection at
various energies in the two materials, also shown in the bottom
panels as an energy distribution integrated over the structure.
At low energies, the carrier collection is overall smaller in aluminum
because of a lower electron-phonon mean free path close to the Fermi level
(in turn because of lighter atoms and higher density of states in Al) \cite{PhononAssisted}.
At higher energies, the mean free path is dominated by
electron-electron scattering which is similar between the two
materials, and the collection probabilities become comparable.
However, at energies above the interband threshold of gold,
accessible by the 390~nm short-axis illumination, aluminum
exhibits much stronger collection of high-energy electrons
compared to gold, because most of the photon energy is
now deposited in the $d$-band holes in gold \cite{NatCom}.
Aluminum wins for high-energy electrons because of a more
favorable initial distribution and comparable carrier transport.

\section{Conclusions and Outlook}
\label{sec:Conclusions}
We have presented a new general theoretical and computational framework, NESSE,
for transport phenomena far from equilibrium and demonstrated its utility
in detail for hot carrier physics in large nanoscale to mesoscale structures.
NESSE hybridizes the best features of phase-space (Boltzmann) and
particle-based (Monte Carlo) methods and allows us to bridge \emph{ab initio}
electronic structure calculations, carrier collision integrals and
electromagnetic simulations to predict the dynamics of photo-excited carriers.
The detailed analysis of scattering mechanisms and transport presented here,
specialized for the case of plasmonic hot carrier dynamics, provides insight
into designing new materials and device motifs which are suitable for carrier transport.
Material design is especially relevant for doped-semiconductor plasmonic materials
where electronic band structures, and hence phase-space for generation
as well as scattering, could be controlled by altering composition.
This work additionally paves the way for geometric design of hot carrier
devices that carefully optimize against and exploit carrier scattering.

In general, NESSE fills a void in theoretical methods to analyze
far-from-equilibrium semi-classical transport phenomena with highly-detailed models
of the scattering processes, such as those involving electrons
and phonons in the hot carrier example treated in detailed here.
It will therefore be invaluable in several areas of physics where
transport of charged particles in strong nonequilibrium is pervasive,
at length scales both microscopic and astronomical.

\section*{Acknowledgments}
This material is based upon work performed at the Joint Center for Artificial Photosynthesis,
a DOE Energy Innovation Hub, supported through the Office of Science
of the U.S. Department of Energy under Award Number DE-SC0004993.
This research used resources of the National Energy Research Scientific Computing Center,
a DOE Office of Science User Facility supported by the Office of Science 
of the U.S. Department of Energy under Contract No. DE-AC02-05CH11231,
as well as the Center for Computational Innovations at Rensselaer Polytechnic Institute.
ASJ thanks the UK Marshall Commission and the US Goldwater Scholarship for financial support.
G.T. acknowledges support from the Swiss National Science Foundation, 
Early Postdoctoral Mobility Fellowship No. P2EZP2-159101.
PN acknowledges start-up funding from the Harvard John A. Paulson School of Engineering and Applied Sciences
and partial support from the Harvard University Center for the Environment (HUCE).
RS acknowledges start-up funding from the Department of Materials Science
and Engineering at Rensselaer Polytechnic Institute.

\bibliographystyle{apsrev4-1}
\makeatletter{} 
\end{document}